\documentclass[a4paper,10pt,twoside,twocolumn]{cpc-hepnp}
\usepackage[pdftex]{hyperref}
\usepackage{CJK,upgreek,fancyhdr}
\usepackage{multicol}
\usepackage{graphicx}
\usepackage{booktabs}
\usepackage{amssymb,bm,mathrsfs,bbm,amscd}
\usepackage[tbtags]{amsmath}
\usepackage{lastpage}
\usepackage{flushend,cuted}

\begin{document}
%\begin{CJK*}{GB}{gbsn}
\begin{CJK*}{GBK}{song}

\title{Anisotropic flow of Pb+Pb $\sqrt{s_{\rm NN}}$ = 5.02 TeV from A Multi-Phase Transport Model\thanks{Supported by National Natural Science
Foundation of China(11135011, 11228513, 11221504)}}

\author{%
Zhao Feng$^{1;1)}$\email{fengzhaoccnu@mails.ccnu.edu.cn}%
\quad Guang-Ming Huang$^{1;2)}$\email{gmhuang@mails.ccnu.edu.cn}%
\quad Feng Liu$^{1;3)}$\email{fliu@mail.ccnu.edu.cn}%
}
\maketitle

\address{%
$^1$ Key Laboratory of Quark and Lepton Physics (MOE) and Institute of Particle Physics,\\ Central China Normal University, Wuhan 430079, China\\
}

\begin{abstract}
Anisotropic flow is an important observable in the study of the Quark-Gluon Plasma that is expected to be formed in heavy-ion collisions. With a multiphase transport (AMPT) model we investigate the elliptic($\emph{v}_{2}$), triangular($\emph{v}_{3}$), and quadrangular($\emph{v}_{4}$) flow of charged particles in Pb+Pb collisions at $\sqrt{s_{\rm NN}}$ = 5.02 TeV. Then We compare our flow results with the published ALICE flow results. We found our AMPT simulated results are consistent with ALICE experimental data.
\end{abstract}

\begin{keyword}
azimuthal anisotropy, anisotropic flow, ALICE Pb+Pb 5.02 TeV, AMPT.
\end{keyword}

\begin{pacs}
12.38.Mh, 25.75.Ld, 25.75.Gz
\end{pacs}
\begin{multicols}{2}

\section{Introduction}
    Ultrarelativistic heavy-ion collisions enable the study of matter at high temperature and pressure where quantum chromodynamics predicts the existence of the quark-gluon
plasma (QGP)~\citep{lab1}. Anisitropic flow, which is caused by the initial asymmetries in the geometry of the system produced in a non-central collision, provides experimental information about the equation of state and the transport properties of the created QGP~\citep{lab2}. Since the transition from normal nuclear matter to the QGP state is expected to occur at extreme values of energy density, elliptic flow has been intensively investigated in some large heavy-ion experimental accelerators like Alternating Gradient Synchrotron(AGS)~\citep{lab3}, Relativistic Heavy-Ion Collider(RHIC)~\citep{lab4,lab5,lab6}, and Large Hadron Collider(LHC)~\citep{lab7,lab8,lab9,lab10}, which lately injected Pb+Pb $\sqrt {{s_{NN}}} $=5.02 TeV beam energy. From the previous studies, azimuthal anisotropy of particle production have contributed significantly to the characterization of the system created in heavy-ion collisions because it is sensitive to the properties of the system at an early time of its evolution. We compare the AMPT model simulate results with string melting mechanism with the ALICE published data, and try to investigate the azimuthal distribution of particles production for different dependencies at LHC energy.

    Anisotropic Flow is characterized by coefficients in the Fourier expansion of the azimuthal dependence of the invariant yield of particles relative to the reaction
plane~\citep{lab11,lab12}:
\setlength\abovedisplayskip{10pt}
\setlength\belowdisplayskip{10pt}
$$E\frac{{{d^3}N}}{{{d^3}p}} = \frac{{{d^2}N}}{{2\pi {p_T}d{p_T}dy}}\left\{ {1 + \sum\limits_{n = 1}^\infty  {2{v_n}\cos \left[ {n(\phi  - {\Psi _R})} \right]} } \right\}\eqno{(1)}$$

    Here ${v_n} = \left\langle {\cos \left[ {n\left( {\phi  - {\Psi _R}} \right)} \right]} \right\rangle$ are coefficients to quantify anisotropic flow. The first coefficient, ${v_1}$, is usually
called directed flow, and the second coefficient, ${v_2}$, is called elliptic flow. In this analysis, we use Q-cumulant method to obtain the anisotropic flow coefficients. Multi-particle correlations can be expressed in terms of flow vector $Q_n$:
\setlength\abovedisplayskip{5pt}
\setlength\belowdisplayskip{5pt}
$${Q_n} \equiv \sum\limits_{i = 1}^M {{e^{in{\phi _i}}}}\eqno{(2)}$$
where M is the number of particles. Then 2-particle and 4-particle azimuthal correlations in one event can be expressed as~\citep{lab13,lab14}:
$$\left\langle 2 \right\rangle  = \frac{{{{\left| {{Q_n}} \right|}^2} - M}}{{M\left( {M - 1} \right)}}\eqno{(3)}$$
$$\left\langle 4 \right\rangle  = \frac{{{{\left| {{Q_n}} \right|}^4} + {{\left| {{Q_{2n}}} \right|}^2} - 2 \cdot {\mathop{\rm Re}\nolimits} \left[ {{Q_{2n}}Q_n^*Q_n^*} \right]}}{{M\left( {M - 1} \right)\left( {M - 2} \right)\left( {M - 3} \right)}}$$
$$ - 2\frac{{2\left( {M - 2} \right) \cdot {{\left| {{Q_n}} \right|}^2} - M\left( {M - 3} \right)}}{{M\left( {M - 1} \right)\left( {M - 2} \right)\left( {M - 3} \right)}}\eqno{(4)}$$
For detectors with uniform acceptance, the $2^{nd}$ order cumulant and $4^{th}$ order cumulant are obtained with:
$${c_n}\left\{ 2 \right\} = \left\langle {\left\langle 2 \right\rangle } \right\rangle \eqno{(5)}$$
$${c_n}\left\{ 4 \right\} = \left\langle {\left\langle 4 \right\rangle } \right\rangle  - 2 \cdot {\left\langle {\left\langle 2 \right\rangle } \right\rangle ^2}\eqno{(6)}$$
Reference flow $\emph{v}_{n}$ estimated from the $2^{nd}$ order cumulant and $4^{th}$ order cumulant are:
$${v_n}\left\{ 2 \right\} = \sqrt {{c_n}\left\{ 2 \right\}} \eqno{(7)}$$
$${v_n}\left\{ 4 \right\} = \sqrt[4]{{ - {c_n}\left\{ 4 \right\}}}\eqno{(8)}$$
For differential cumulant, we use p-vector and q-vector derived from Eq.(2):
$${p_n} = \sum\limits_{i = 1}^{{m_p}} {{e^{in{\phi _i}}}} \eqno{(9)}$$
$${q_n} = \sum\limits_{i = 1}^{{m_q}} {{e^{in{\phi _i}}}} \eqno{(10)}$$
Here $\emph{m}_{p}$ is the total number of particles labeled as POIs(Particle Of Interest), $\emph{m}_{q}$ is the total number of particles tagged both as RFP(Reference Particle) and POI. And the single-event average differential cumulant goes to:
$$\left\langle {2'} \right\rangle  = \frac{{{p_n}Q_n^* - {m_q}}}{{{m_p}M - {m_q}}}\eqno{(11)}$$
\[\begin{array}{l}
\left\langle {4'} \right\rangle  = \left[ \begin{array}{l}
{p_n}{Q_n}Q_n^*Q_n^* - {q_{2n}}Q_n^*Q_n^* - {p_n}{Q_n}Q_{2n}^* - 2 \cdot M{p_n}Q_n^*\\
 - 2 \cdot {m_q}{\left| {{Q_n}} \right|^2} + 7 \cdot {q_n}Q_n^* - {Q_n}q_n^* + {q_{2n}}Q_{2n}^*\\
 + 2 \cdot {p_n}Q_n^* + 2 \cdot {m_q}M - 6 \cdot {m_q}
\end{array} \right]\\
\begin{array}{*{20}{c}}
{}&{}
\end{array}/\left[ {\left( {{m_p}M - 3{m_q}} \right)\left( {M - 1} \right)\left( {M - 2} \right)} \right]
\end{array}\eqno{(12)}\]
For detectors with uniform azimuthal acceptance the differential $2^{nd}$  order cumulant and $4^{th}$  order cumulant are given by:
\[{d_n}\left\{ 2 \right\} = \left\langle {\left\langle {2'} \right\rangle } \right\rangle \eqno{(13)}\]
\[{d_n}\left\{ 4 \right\} = \left\langle {\left\langle {4'} \right\rangle } \right\rangle  - 2\left\langle {\left\langle {2'} \right\rangle } \right\rangle \left\langle {\left\langle 2 \right\rangle } \right\rangle \eqno{(14)}\]
Finally:
\[v{'_n}\left\{ 2 \right\} = \frac{{{d_n}\left\{ 2 \right\}}}{{\sqrt {{c_n}\left\{ 2 \right\}} }}\eqno{(15)}\]
\[v{'_n}\left\{ 4 \right\} = \frac{{{d_n}\left\{ 4 \right\}}}{{{{\left( { - {c_n}\left\{ 4 \right\}} \right)}^{3/4}}}}\eqno{(16)}\]
However, non-flow effects which are produced from resonance decays and jets should be reduced in the correlation, thus a $\eta$ gap need to be appied for 2-particle correlation, and the Eq.(7) and Eq.(15) are changed to~\citep{lab15}:
\[\left\langle 2 \right\rangle  = \frac{{Q{^A_n} \cdot Q{^B_n}^*}}{{M{_A}M{_B}}}\eqno{(17)}\]
\[\left\langle {2'} \right\rangle  = \frac{{p{^A_n} \cdot Q{^B_n}^*}}{{{m_{p,A}}M_B}}\eqno{(18)}\]
While $Q{^A_n}$ and $Q{^B_n}$ mean the 2 Q-vectors of left and right side of the gap, same for the $p{^A_n}$.

    In this analysis, we use the events simulated from a multiphase transport(AMPT) model~\citep{lab16} to obtain anisotropic flow coefficient. The AMPT model is constructed to
describe nuclear collisions ranging from p+A to A+A systems at center-of-mass energies from about $\sqrt {{s_{NN}}} $ = 5 GeV up to 5500 GeV at LHC, where strings and minijets dominate the initial energy production and effects from final-state interactions are important. It consists of four main components: the initial conditions, partonic interactions, the conversion from the partonic to the hadronic matter, and hadronic interactions. The initial conditions are generated by the heavy-ion jet interaction generator (HIJING) model, the strings are converted into partons and the next stage, which models the interactions between all the partons, is based on ZPC(Zhang's parton cascade~\citep{lab17}).  In ZPC, the default value of the cross section is 3 mb. The transition from partonic to hadronic matter is modeled by a simple coalescence model, which combines two quarks into mesons and three quarks into baryons. And the dynamics of the subsequent hadronic matter is described by a hadronic cascade, which is based on the ART model. We used AMPT version v2.26t5 with Lund parameter a=0.30, b=0.15/$GeV^2$ in this analysis. The anisotropic flow for Pb+Pb $\sqrt {{s_{NN}}} $=5.02 TeV from AMPT model has been generally investigated~\citep{lab18}, and in this anaysis we would like to apply the ALICE TPC cut specifically and compare the simulated results with ALICE newly published results~\citep{lab10}.

\section{Results and discussions}
\begin{center}
\includegraphics[width=8.5cm]{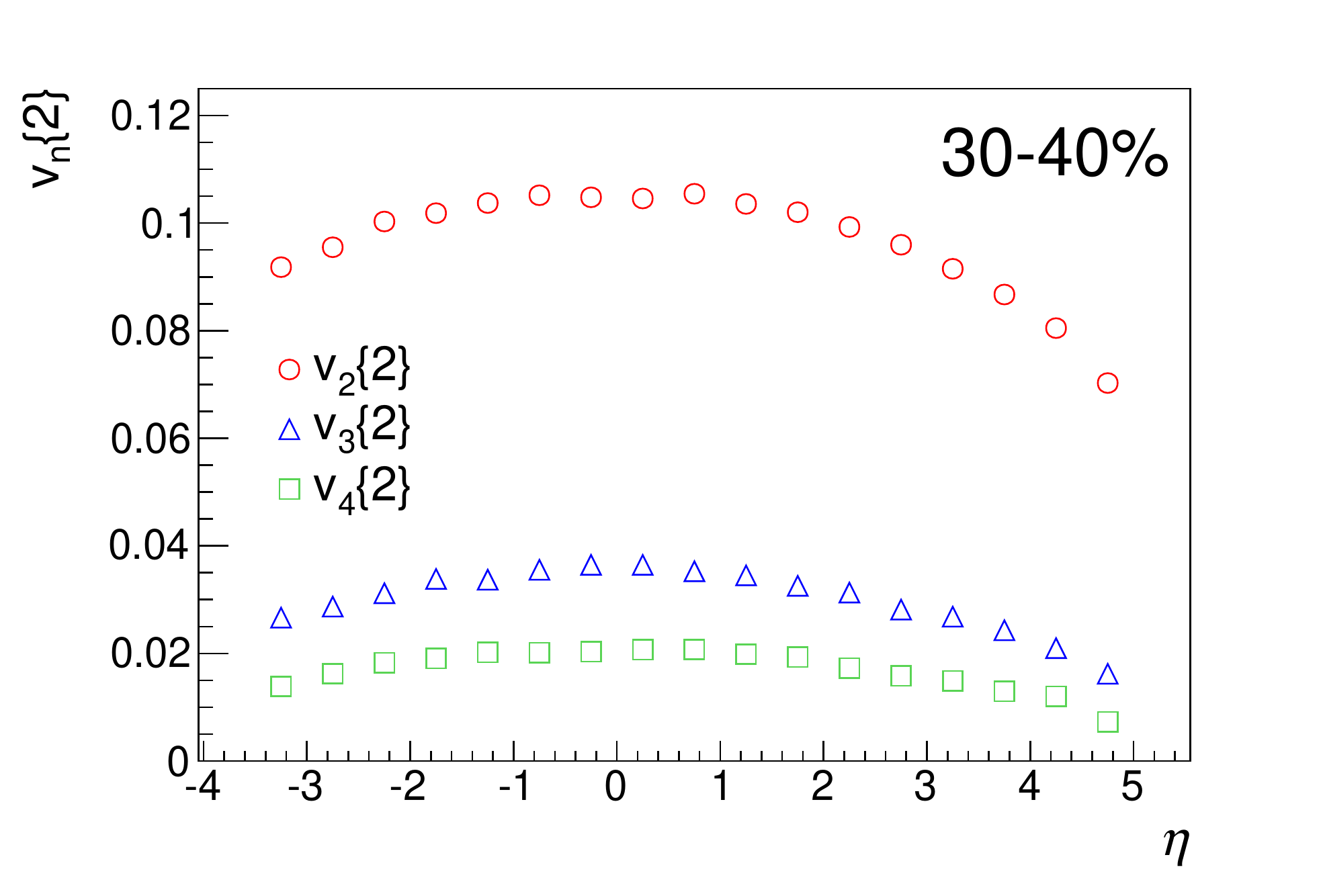}
\figcaption{\label{fig1} ${v_n\{2\}}$ as a function of pseudorapidity in $ - 3.5 < \eta  < 5$ range for centrality 30-40\%.}
\end{center}

    In 2015, LHC launched Pb+Pb collisions at $\sqrt {{s_{NN}}} $=5.02 TeV. In this analysis, we use the results obtained from AMPT simulated minimal bias events to compare
with ALICE experimental data in Run 2. We use the transverse momentum range 0.2$<{p_T}<$5.0 GeV, pseudorapidity range $ - 0.8 < \eta  < 0.8$, in order to keep the same $\eta$ and ${p_T}$ cuts with ALICE data. In this analysis, we use 150k AMPT simulated minimal-bias Pb+Pb $\sqrt {{s_{NN}}} $=5.02 TeV events to extract flow coefficients, to make sure all simulated results are in fairly low uncertainty.

   The psuedorapidity($\eta$) dependence of ${v_2}$, ${v_3}$, ${v_4}$ for 30-40\% most central collision are presented in Fig.\ref{fig1}. We otained this result using the same Q-cumulant method with ALICE Pb+Pb $\sqrt {{s_{NN}}} $=2.76 TeV experimental result~\citep{lab19}. We can see that ${v_2}$ is obviously larger than
${v_3}$, ${v_4}$. The results show the distribution for all of these 3 harmonics are flat in middle rapidity region($ - 0.8 < \eta  < 0.8$). So we could integrate this dimension to obtain anisotropic flow of other physical quantity.

\begin{center}
\includegraphics[width=8.5cm]{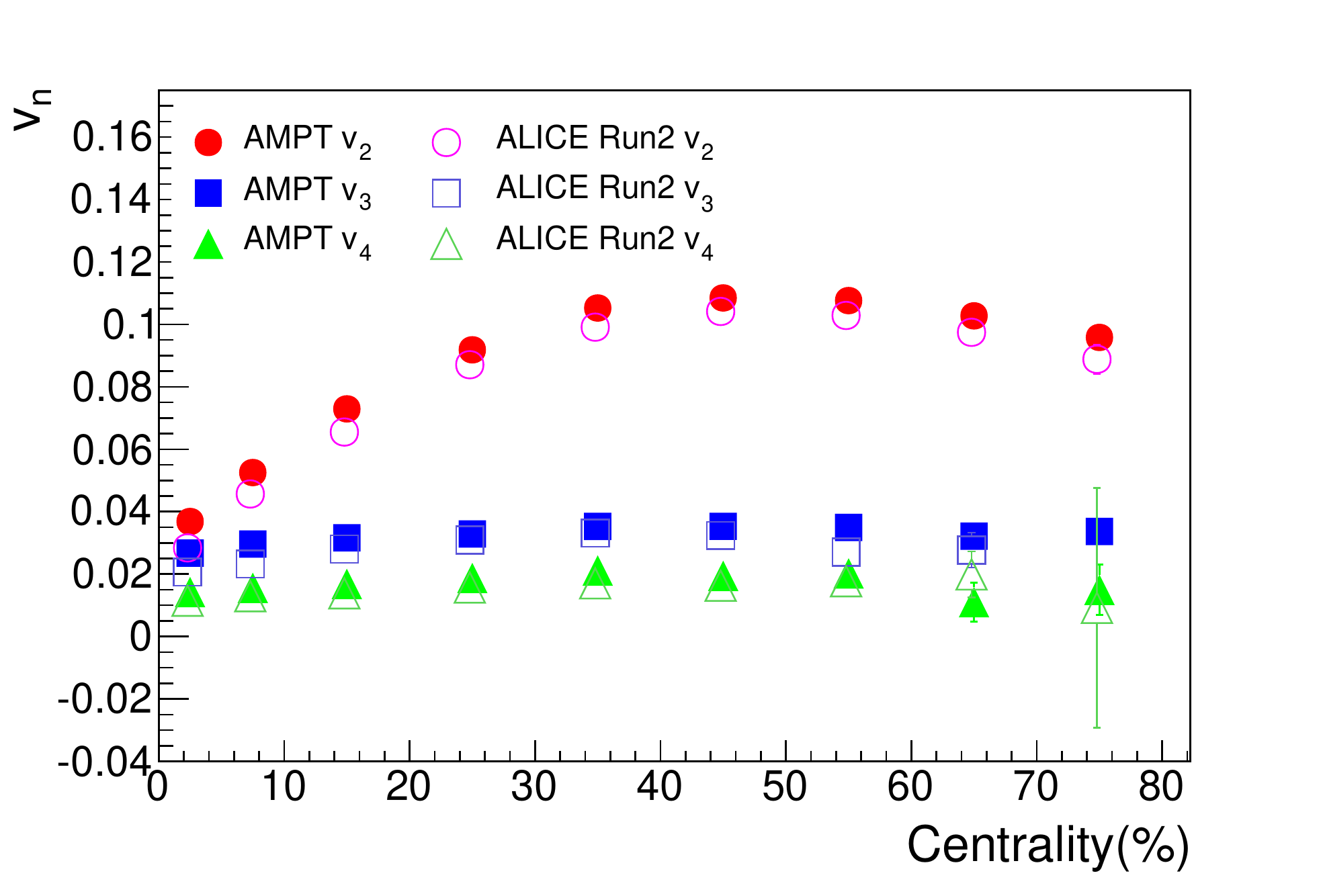}
\figcaption{\label{fig2}Anisotropy flow ${v_n}$ as a function of event centrality, using two-particle cumulant method with $\left| {\Delta \eta } \right| > 1$.  Solid markers are for AMPT simulated results while open markers are for ALICE Pb+Pb 5.02 TeV experimental data. }
\end{center}

    In Fig.\ref{fig2} a clear centrality dependence of ${v_2}$ is observed, increasing from central to middle-central collisions, saturating in 40-50\% centrality class, and then
decreasing as the interactions during the system evolution diluted. For ${v_3}$ and ${v_4}$, the centrality dependence is relative weaker, compared to ${v_2}$. It is also seen in Fig.\ref{fig2} that AMPT calculations reproduce successfully the centrality dependence of ${v_n}$, slightly overestimate the data.

\begin{figure*}
\begin{center}
\includegraphics[width=8.5cm]{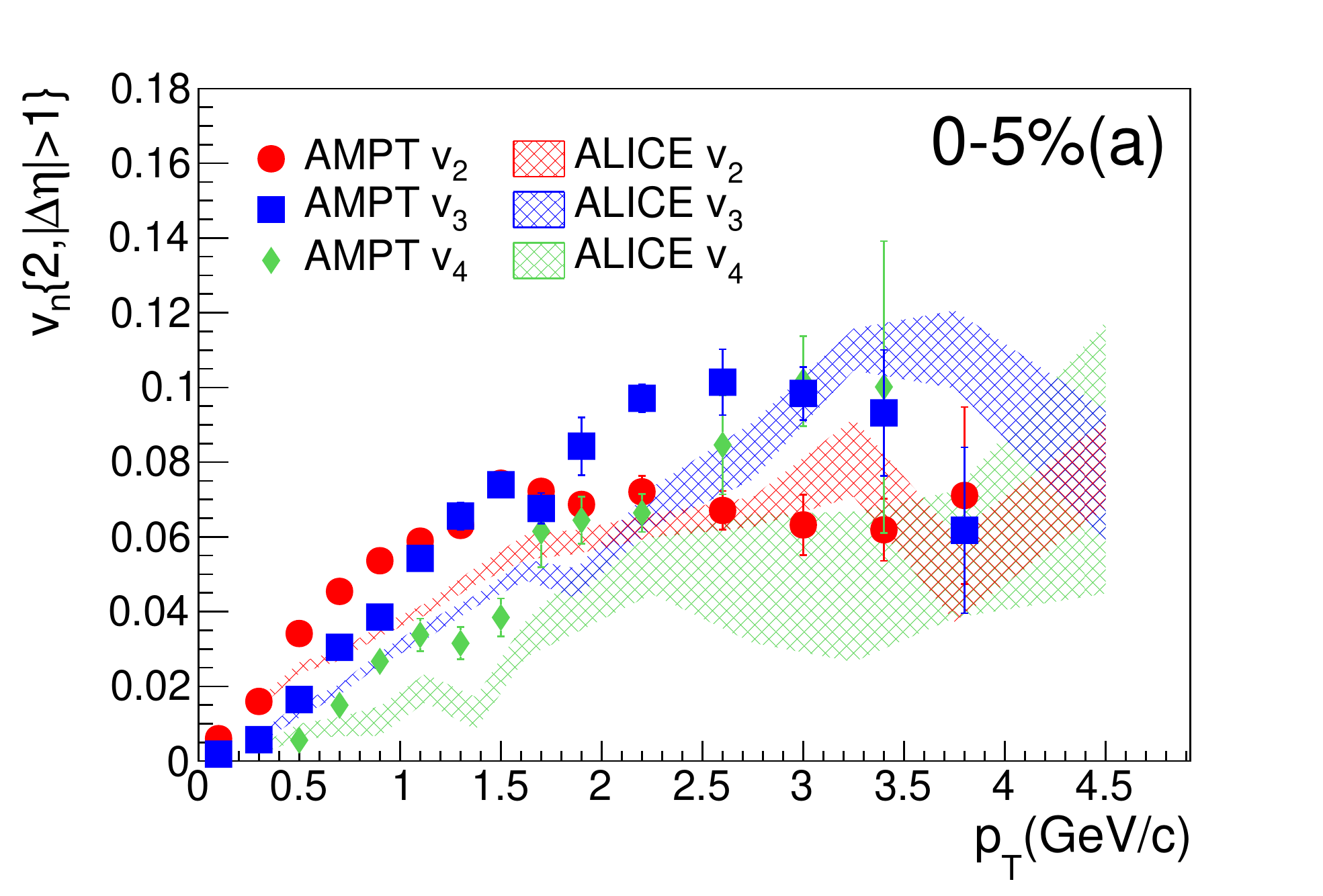}
\includegraphics[width=8.5cm]{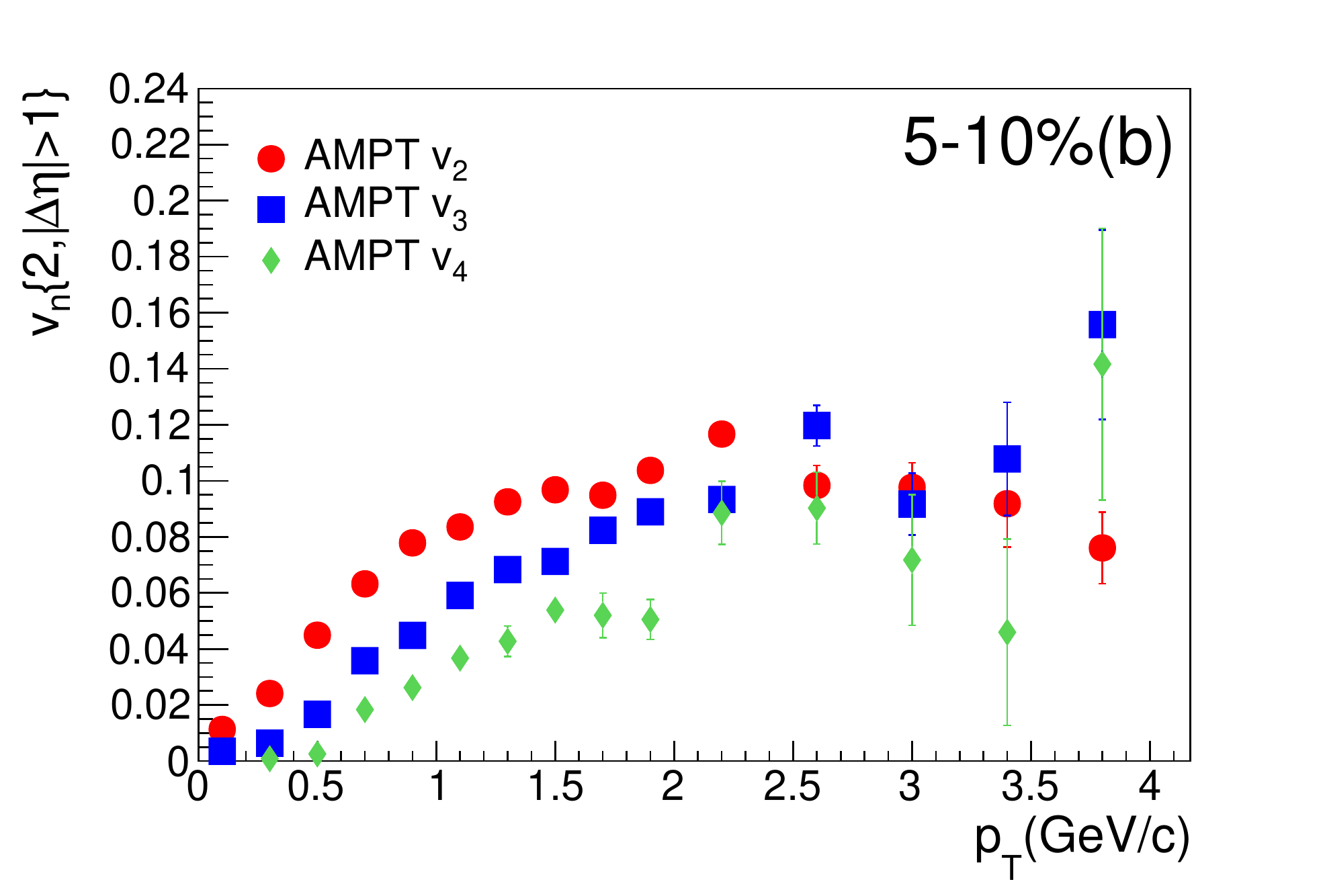}
\includegraphics[width=8.5cm]{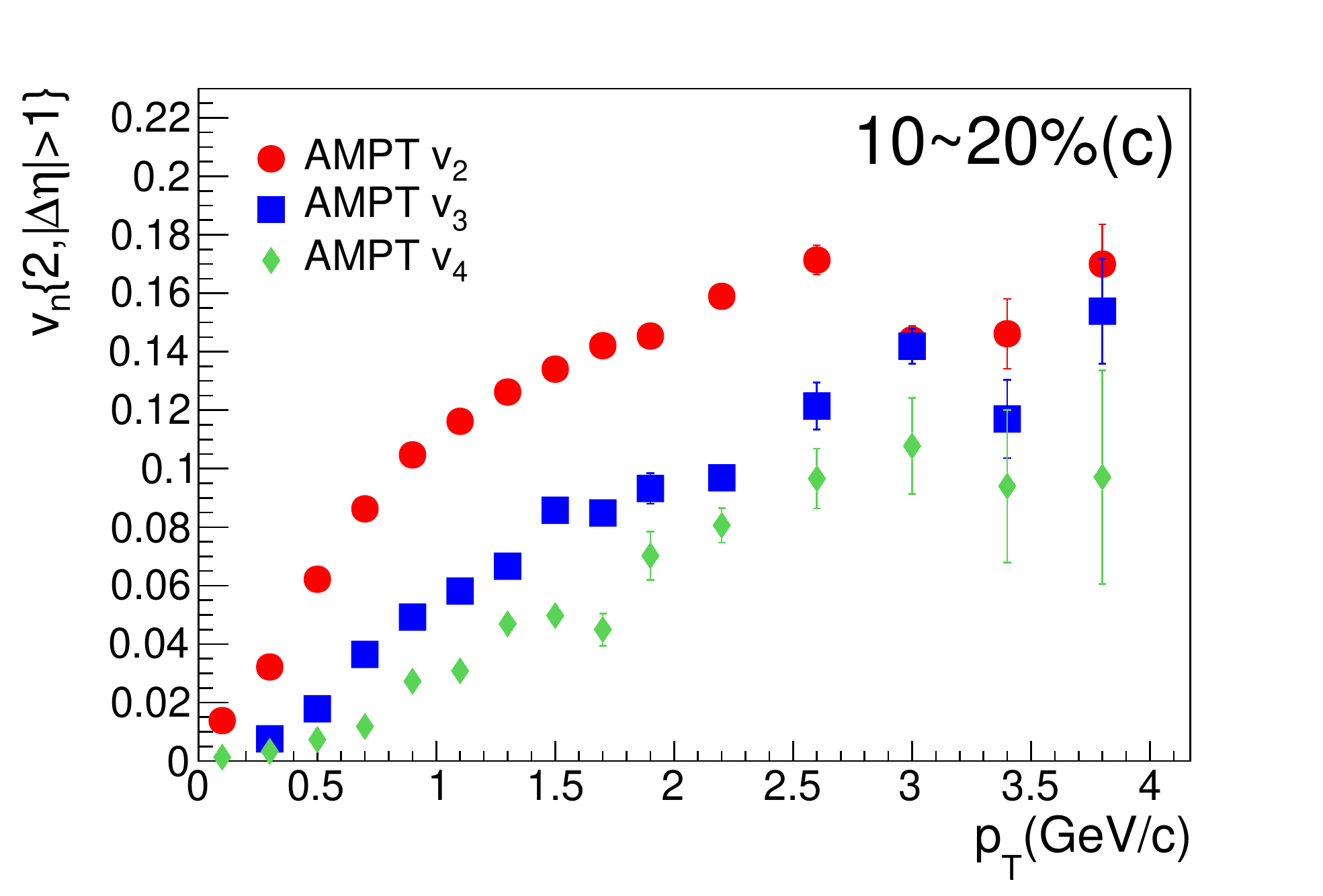}
\includegraphics[width=8.5cm]{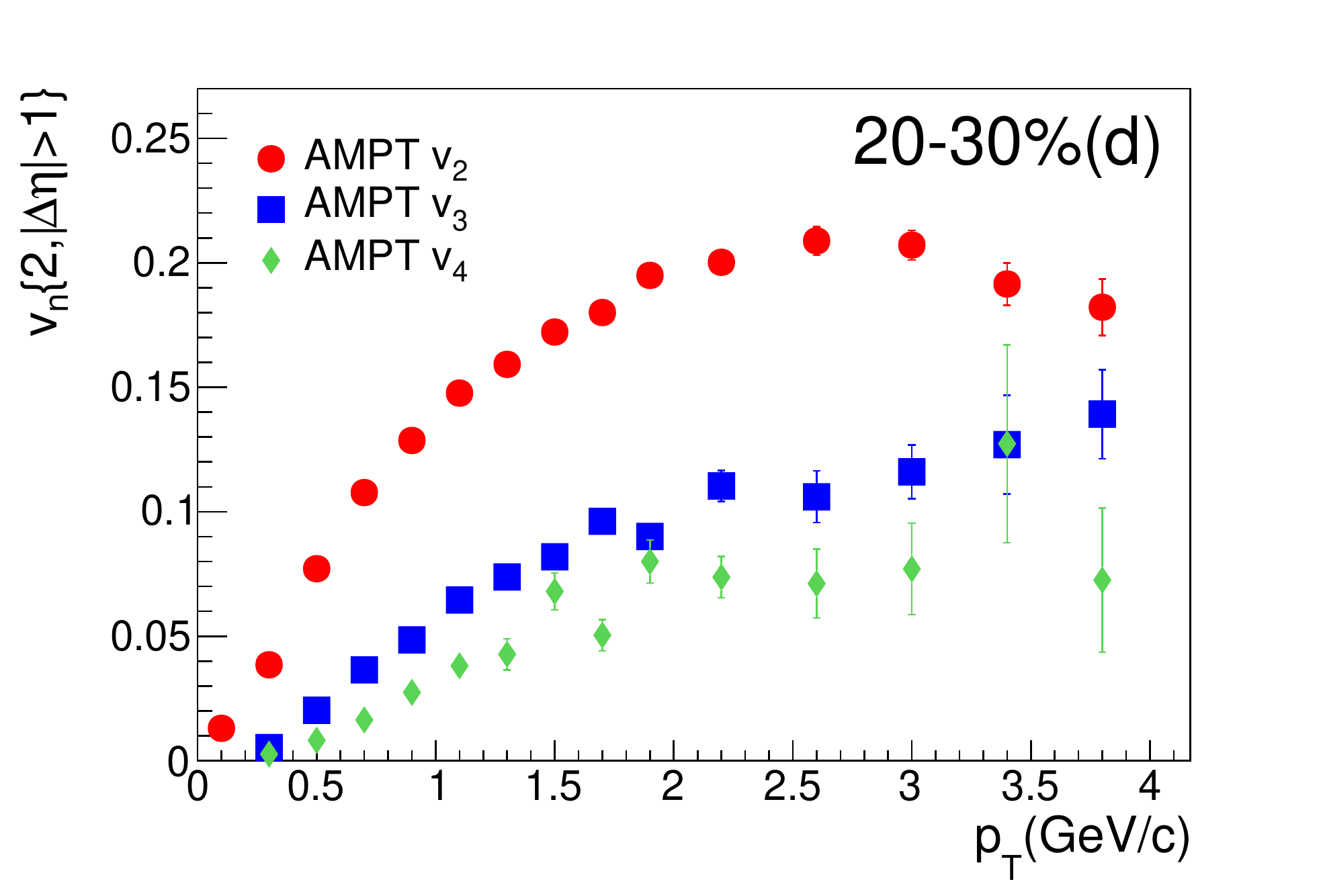}
\includegraphics[width=8.5cm]{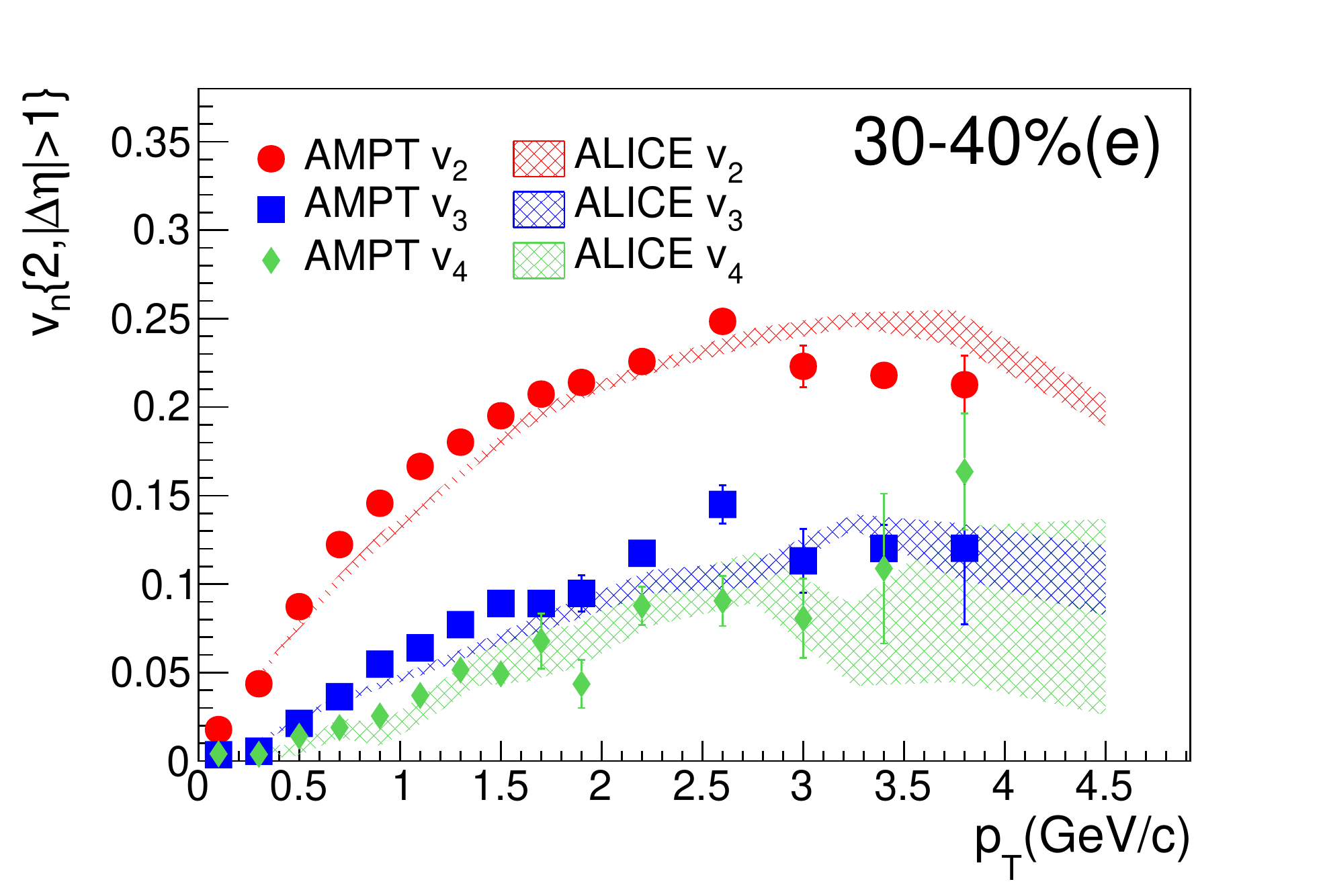}
\includegraphics[width=8.5cm]{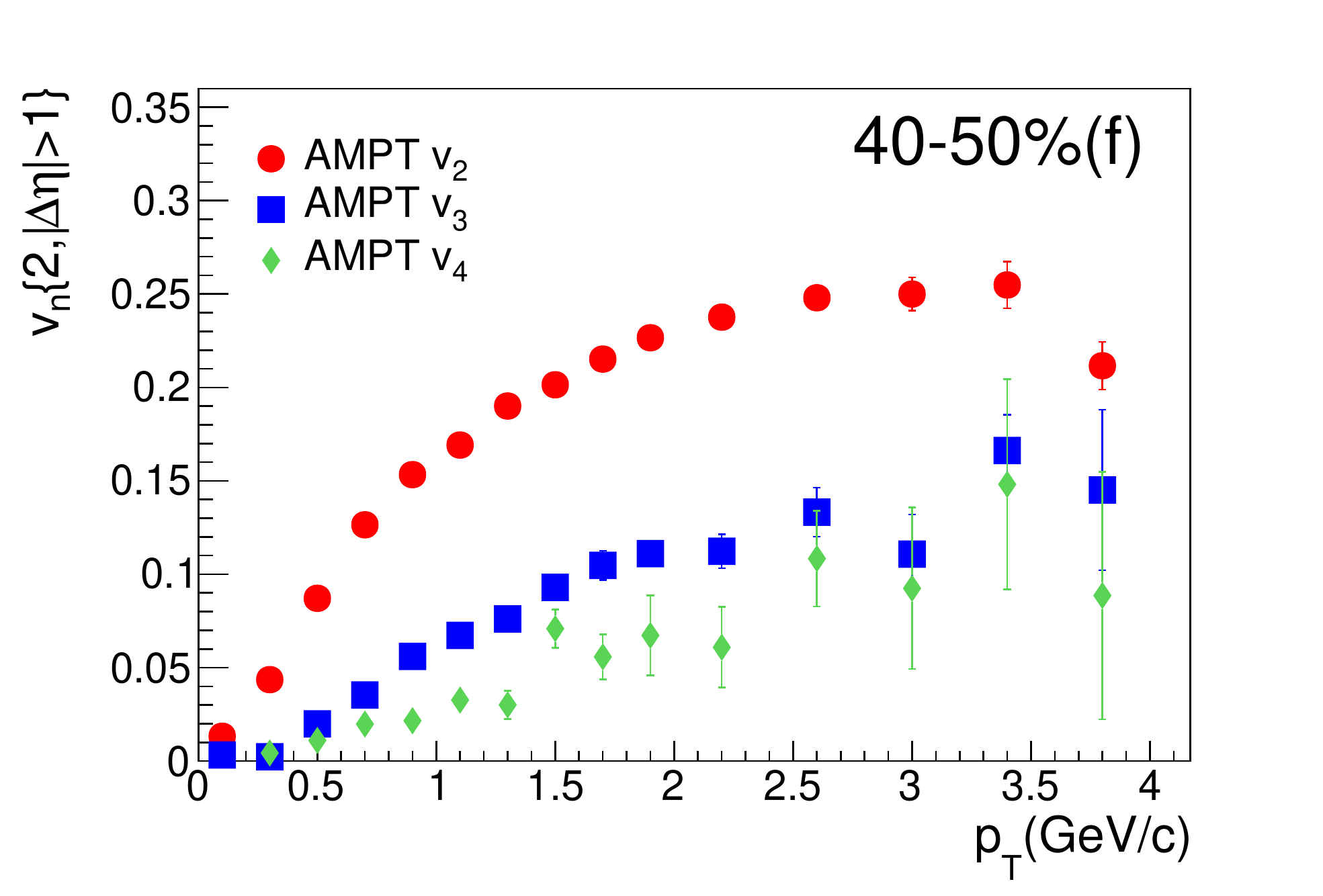}
\figcaption{\label{fig3}Anisotropy flow ${v_n}$ as a function of transverse momentum from 0-5\% to 40-50\%, using two-particle cumulant method with $\left| {\Delta \eta } \right| > 1$, compared with ALICE published 0-5\% and 30-40\% experimental results. Solid dots are for AMPT results and shadow grids are for ALICE Pb+Pb 5.02TeV data with grid height corresponds to uncertainties.}
\end{center}
\end{figure*}

    Fig.\ref{fig3} shows the transverse momentum dependence of ${v_2}$\{{2,${\Delta \eta }>1$}\}. We can see that from 0-5\% to 40-50\%, the anisotropic flow signals are
increasing as centrality increases, while the difference between ${v_2}$ and ${v_3}$, ${v_4}$  is getting bigger too. For 0-5\% most central collisions(see Fig.\ref{fig3}(a)), AMPT qualitatively reproduce the pt-differential anisotropic flow, it works better in more peripheral collisions. For 30-40\%(see Fig.\ref{fig3}(e)), the situation seems better. The two results are well consistent except for few high momentum points($>$3.5GeV/c).

\begin{center}
\includegraphics[width=8.5cm]{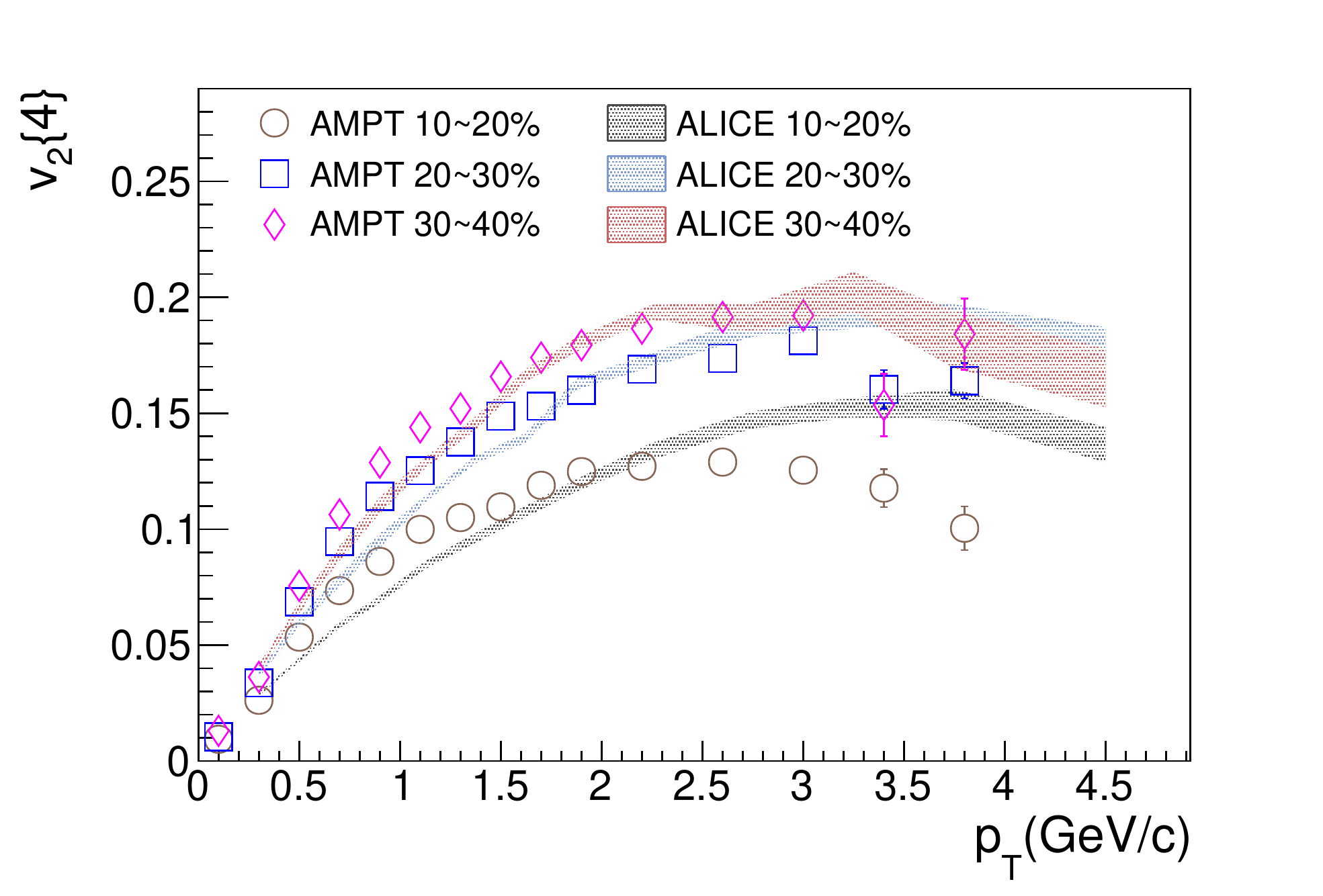}
\figcaption{\label{fig4}Elliptic flow ${v_2\{4\}}$ as a function of transverse momentum for 10-20\%, 20-30\%, 30-40\%, no $\eta$ gap applied.}
\end{center}

    Fig.\ref{fig4} represents the ${v_2}$ as a function of transverse momentum using 4-particles cumulant method.  We can see AMPT calculations correctly reproduce the
pt-depdence of anisotropic flow. The agreement between data and AMPT calculations seems better in more peripheral collisions, as we also observed in Fig.\ref{fig3}.

\section{Conclusions}
    We did systematic studies of the harmonic flow in Pb+Pb collisions at center of mass energy of 5.02 TeV with a multi-phase transport model. Centrality dependence of
anisotropic flow has been presented, as well as the comparisons to the published measurements from ALICE. And for different centrality classes, anisotropy flow on transverse momentum's dependence has also been compared. For centrality dependence our AMPT results are consistent with the experimental data, higher harmonics $v_3$ and $v_4$ which come from the flow fluctuation are showing the same flatten distribution as ALICE result. For $p_T$ dependence the AMPT model could quite well reproduce the ${v_2}$ of experimental data. We can see from the comparisons that the string melting version of the AMPT model can describe the qualitative features of flow distribution, and it can reproduce the experimental data quantitatively.

\section{Acknowledgement}
    This work is supported by the MOST of China under 973 Grant 2015CB856901, the National Natural Science Foundation of China under grants No.11135011, 11228513,
11221504. We also thank Dr.You Zhou and Prof.Ziwei Lin for their discussions and suggestions along with the paper.
\end{multicols}
\vspace{-1mm}
\centerline{\rule{80mm}{0.1pt}}
\vspace{2mm}

\begin{multicols}{2}

\end{multicols}
\clearpage
\end{CJK*}
\end{document}